\documentclass[%
reprint,
superscriptaddress,
showpacs,preprintnumbers,
amsmath,amssymb,
aps,
]{revtex4-1}

\usepackage{color}
\usepackage{graphicx}
\usepackage{dcolumn}
\usepackage{bm}

\begin {document}

\title{Phase Diagram in Stored-Energy-Driven L\'evy Flight}

\author{Takuma Akimoto}
\email{akimoto@z8.keio.jp}
\affiliation{%
  Department of Mechanical Engineering, Keio University, Yokohama, 223-8522, Japan
}%

\author{Tomoshige Miyaguchi}
\email{tmiyaguchi@naruto-u.ac.jp}
\affiliation{%
  Department of Mathematics Education, Naruto University of Education, Tokushima 772-8502, Japan
}%


\date{\today}

\begin{abstract}
Phase diagram based on the mean square displacement (MSD) 
 and the distribution of diffusion coefficients of the time-averaged 
 MSD for the stored-energy-driven L\'evy flight (SEDLF) is presented. 
 In the SEDLF, a random walker cannot move while storing energy, and 
 it jumps by the stored energy. The SEDLF shows a whole spectrum of 
 anomalous diffusions including subdiffusion and superdiffusion, depending 
 on the coupling parameter between storing time (trapping time) and stored 
 energy. This stochastic process can be investigated analytically with the aid 
 of renewal theory. Here, we consider two different renewal processes, i.e., 
 ordinary renewal process and equilibrium renewal process, when the mean 
 trapping time does not diverge. We analytically show the phase diagram 
 according to the coupling parameter and the power exponent in the 
 trapping-time distribution.   In particular, we find that distributional 
 behavior of time-averaged MSD intrinsically appears in superdiffusive as 
 well as normal diffusive regime even when the mean trapping time does not diverge.
\end{abstract}

\maketitle

\maketitle

\section {Introduction}

In {\color{black}normal} diffusion processes, the diffusivity can be characterized by the diffusion coefficient in the mean square displacement (MSD). 
However, many diffusion processes in nature show anomalous diffusion; 
that is, the MSD does not grow linearly with time but follows a sublinear or superlinear growth with time,   
\begin{equation}
\langle x_t^2 \rangle \propto t^\beta \quad(\beta\ne 1), 
\end{equation} 
{\color{black}where $x_t$ is a position in one-dimensional coordinate, $t$ is time, and $\langle \cdot \rangle$ means an ensemble average.} 
In particular, anomalous diffusion in biological systems has been found by single-particle tracking experiments 
\cite{Caspi2000,Golding2006,Graneli2006,Weigel2011,Jeon2011,Weber2012,Tabei2013}.  
 Thus, the power-law exponent $\beta$ is one of the most important quantities {\color{black}characterizing}
  the underlying diffusion process. Especially, anomalous diffusion with $\beta <1$ is called 
 subdiffusion and that with $\beta>1$ is called superdiffusion.  
  
  Although anomalous diffusion can be characterized by the power-law exponent $\beta$ in the MSD, 
  the exponent cannot reveal the underlying physical nature in itself. This is because 
  the same power-law exponent in the MSD does not imply that the physical mechanism in the anomalous diffusion 
  is also the same. Therefore, clarifying the origin of anomalous diffusion is an important subject, and 
  many researches on this issue have been conducted extensively \cite{Magdziarz2009,Tejedor2010a,Kepten2011,Magdziarz2011,Meroz2013}.
  {\color{black}One of the key properties characterizing anomalous diffusion is ergodicity, i.e., time-averaged observables being equal to 
  a constant (the ensemble average). In some experiments \cite{Golding2006,Graneli2006,Weigel2011,Jeon2011,Tabei2013}, (generalized) 
  diffusion coefficients for time-averaged MSDs show large fluctuations, suggesting that ergodicity breaks.
  }
  
  In stochastic models of anomalous diffusion, continuous-time random walk (CTRW) shows a prominent feature called 
  {\it distributional ergodicity} \cite{Lubelski2008,He2008,Miyaguchi2011a,Miyaguchi2011}; that is, the time-averaged observables 
  obtained from single trajectories do not converge to a constant but the distribution of such time-averaged observables 
   converges to a universal distribution ({\it convergence in distribution}). 
  More precisely, the distribution of the
time-averaged MSD (TAMSD), which is defined by
\begin{equation}
  \overline{\delta^2 (\Delta; t)} \equiv \frac{1}{t-\Delta} \int_0^{t-\Delta} (x_{t'+\Delta}-x_{t'})^2 dt', 
  \label{tamsd}
\end{equation}
converges to the Mittag-Leffler distribution of order $\alpha$
\cite{He2008,Akimoto2010}.  This statement can be represented by
\begin{equation}
  \frac{\overline{\delta^2 (\Delta; t)}}{\langle \overline{\delta^2 (\Delta; t)}\rangle}
  \Rightarrow M_\alpha\quad {\rm as}~t\to \infty,
\end{equation}
for a fixed $\Delta$ {\color{black}($\ll t$)}, where $M_\alpha$ is a random variable with the
Mittag-Leffler distribution of order $\alpha$. We note that the diffusion coefficients in the TAMSDs are also distributed 
according to the Mittag-Leffler distribution because the TAMSD shows normal diffusion, i.e, 
$\overline{\delta^2 (\Delta; t)} \simeq D_t \Delta$ \cite{Miyaguchi2011a,Miyaguchi2011}, {\color{black} where we refer to $D_t$ as 
the diffusion coefficient. In stochastic models,}
   such distributional behavior originates from 
  the divergent mean trapping time. {\color{black}In  diffusion in a random energy landscape such as a trap model \cite{bouchaud90}, 
  the trapping-time distribution follows a power law, $w(\tau) \propto \tau^{-1-\alpha}$, if heights of the energy barrier are distributed according 
  to the exponential distribution.}
  The exponent $\alpha$ smaller than 1 implies a divergence of the mean. {\color{black}In dynamical systems, this divergent mean brings an infinite 
  measure \cite{Akimoto2010a}. Thus, }
  such distributional behavior of time-averaged observables is also shown in {\it infinite ergodic theory} \cite{Akimoto2010,Aaronson1997}. 
  {\color{black}Moreover,  large fluctuations of time-averaged observables, related to distributional ergodicity, have been observed in biological experiments 
  \cite{Golding2006,Graneli2006,Weigel2011,Jeon2011,Tabei2013} as well as quantum dot experiments \cite{Brok2003,Stefani2009}.}
    
  Recently, we have shown that the distribution of the diffusion coefficients of the TAMSDs
  in the stored-energy-driven L\'evy flight (SEDLF) is different from that in CTRW \cite{Akimoto2013a}. 
   {\color{black}The SEDLF is a CTRW with jump lengths correlated with
   trapping times \cite{Klafter1987,Magdziarz2012,Liu2013}.  One of the
   most typical examples for such a correlated motion can be observed in
   L\'{e}vy walk \cite{Shlesinger1987}. However, L\'{e}vy walk and SEDLF
   are completely different stochastic processes in that a random walker
   cannot move while it is traped in SEDLF, whereas it can move with
   constant velocity in L\'{e}vy walk.  In other words, in SEDLF, a random
   walker does not move while storing a sort of energy, and it jumps using
   the stored energy (see Fig.~1).  When the trapping-time distribution
   follows a power-law, the jump length distribution also follows a
   power-law, the same as in L\'evy flight.  Since we consider a power-law
   trapping-time distribution, we refer to our model as the
   stored-energy-driven L\'evy flight.  We note that the MSD does not
   diverge in SEDLF, whereas it always diverges in L\'evy flight.  Although
   the ensemble-averaged MSDs show subdiffusion as well as superdiffusion,
   the TAMSDs always increase linearly with time in SEDLF
   \cite{Akimoto2013a}. This behavior is completely different from that in
   L\'{e}vy walk \cite{Akimoto2012,Froemberg2013}.  Moreover,}
   the distribution of the TAMSD with a fixed $\Delta$ converge to a time-independent 
  distribution which is not the same as  {\color{black} a universal distribution in CTRW (Mittag-Leffler distribution):} 
  \begin{equation}
  \frac{\overline{\delta^2 (\Delta; t)}}{\langle \overline{\delta^2 (\Delta; t)}\rangle}
   \Rightarrow Y_{\alpha,\gamma} \quad {\rm as}~t\to \infty,
  \end{equation}
  where $Y_{\alpha,\gamma}$ is a random variable and $\gamma$  is 
  the coupling parameter between trapping time and jump length. We note that such coupling effects 
   become physically important in {\color{black}turbulent diffusion \cite{Shlesinger1987}, diffusion of cold atoms \cite{Barkai2014}, and }
   nonthermal systems such as cells \cite{Caspi2000,Weber2012}. {\color{black}In such enhanced diffusions, it has been known that the coupling between 
   jump lengths and waiting times follows a power-law fashion like SEDLF \cite{Shlesinger1987,Barkai2014}, although a particle is always moving, 
   which is different from SEDLF. Furthermore,
    it will also be important 
   in complex systems such as finance \cite{Meerschaert2006} and earthquakes \cite{Corral2006,Lippiello2013} because 
   jump lengths are correlated with the waiting times in such systems. }

   {\color{black}In terms of an ensemble average}, SEDLF exhibits a whole spectrum
    of diffusion: sub-, normal-, and super-diffusion, depending on the coupling  parameter
    \cite{Akimoto2013a,Klafter1987,Magdziarz2012,Liu2013}.  
    Because distributional behavior of the time-averaged 
    observables such as the diffusion coefficients in SEDLF is different from that in CTRW, it is important to construct 
    a phase diagram in terms of the power-law exponent of the MSD as well as the form of the distribution function of 
    the TAMSD. 
    Here, we provide the phase diagram for the whole parameters range in  SEDLF.

\section {Model}

  SEDLF is a cumulative process, which is a generalization of a renewal process \cite{Cox}. 
  Equivalently, SEDLF is a CTRW with a non-separable
  joint probability of trapping time and jump length. Therefore, the SEDLF can be 
  defined through the joint probability density function (PDF) $\psi (x, t)$,
  where $\psi (x, t)dx dt$ is the probability that a random walker jumps with
  length $[x, x+dx)$ just after it is trapped for period $[t, t+dt)$ since its previous
  jump \cite{bouchaud90,Shlesinger1982}.  Here, we consider 
 the following joint PDF 
 \begin{equation}
  \label{e.joint-prob}
  \psi(x,t) = w(t) \frac {\delta (x-t^{\gamma}) + \delta (x+t^{\gamma})}{2},
  \end{equation}
  where $w(t)$ is the PDF of trapping times and {\color{black}$0 \leq \gamma \leq 1$} is a coupling strength. 
  {\color{black}This kind of coupling has been introduced in \cite{Klafter1987,Shlesinger1982}.}
  The SEDLF with $\gamma=0$ is just a separable CTRW. 
  In addition, we consider that the PDF of trapping times follows a power law:
  \begin{equation}
  \label{e.pdf-trap-jump}
  w(t) \simeq \frac {c_0}{t^{1+\alpha}},
  \end{equation}
  as $t\to \infty$. Here, $\alpha \in (0,2)$ is the stable index, a constant
  $c_0$ is defined by $c_0 = {c}/{|\Gamma(-\alpha)|}$ with a scale factor $c$. 
  We note that the mean trapping time diverges for $\alpha \leq 1$. 
  For $\gamma>0$, the PDF of jump length also 
  follows a power law:
  \begin{equation}
    \label{e.pdf-jump-length}
    l(x) = \int_{0}^{\infty} \psi (x,t) dt  =
    \frac {|x|^{\frac {1}{\gamma}-1}}{2\gamma}w\left(|x|^{\frac
      {1}{\gamma}}\right)
    \simeq \frac {c_0}{2\gamma}\frac {1}{|x|^{1+ {\alpha} / {\gamma}}}.
  \end{equation}
  Thus, the second moment of the jump length diverges for $2\gamma \geq \alpha$.
  Because L\'evy flight also has a power law distribution of jump
  length, we call this random walk the stored-energy-driven L\'evy flight.
  In numerical simulations, we set the PDF of the trapping time as 
  $w(t) = \alpha t^{-1-\alpha}$ $(t\geq 1)$. Thus, the jump length PDF is given by {\color{black}$l(x) =
    \alpha/(2\gamma |x|^{1+\alpha/\gamma})$} from
    Eq.~(\ref{e.pdf-jump-length}), and  $\langle l^2 \rangle =
    \alpha/(\alpha-2\gamma)$ for $2\gamma < \alpha$.  
    
\begin{figure}
 \includegraphics[width=.9\linewidth, angle=0]{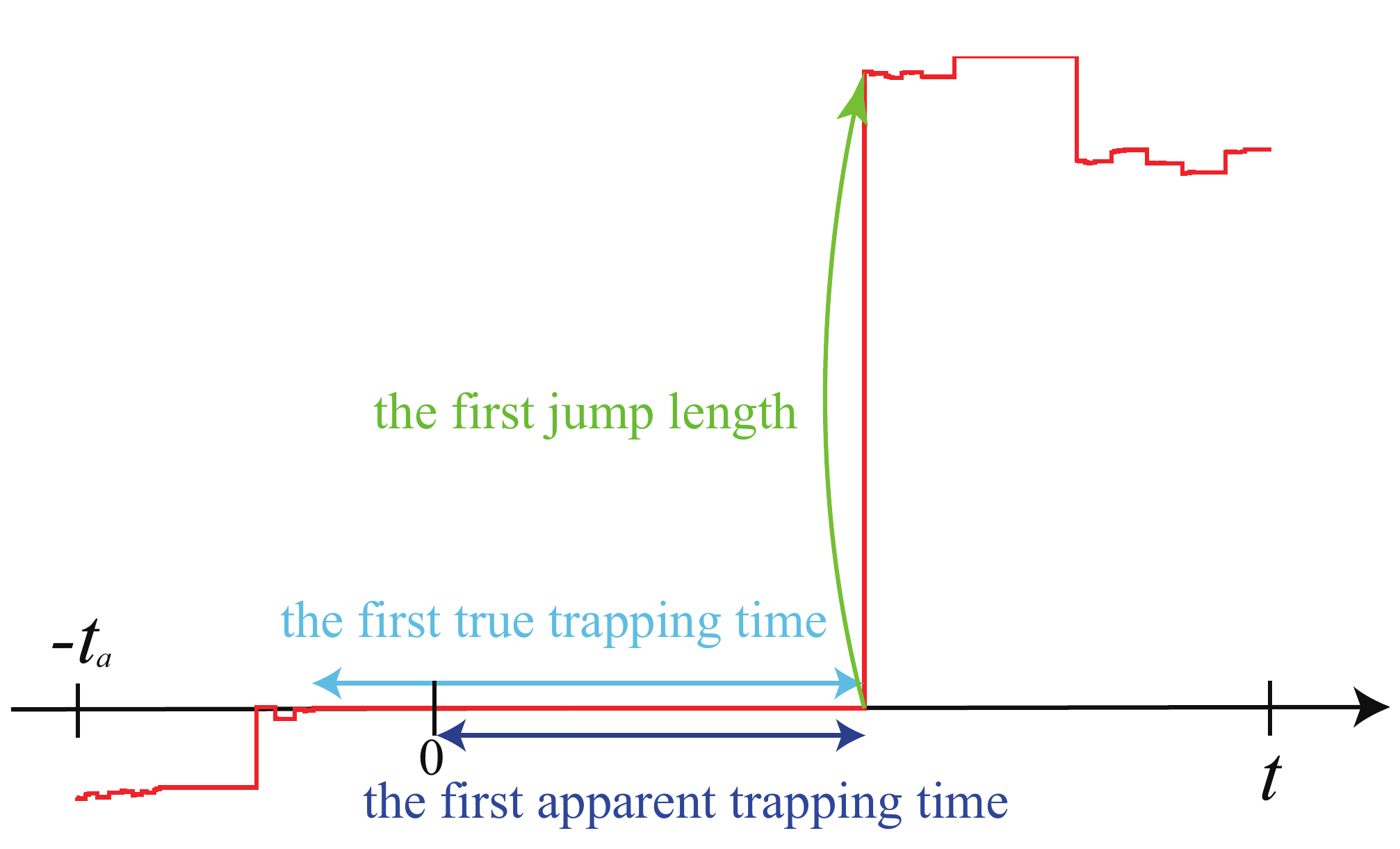}
  \caption{ Trajectory of SEDLF in an equilibrium renewal process. 
  A measurement starts at $t=0$ while the process starts at $-t_a$. 
      An equilibrium process means a process with $t_a \to \infty$ if there exist an equilibrium distribution 
      for the first apparent time and the second moment of the first jump length.}
  \label{traj_eq}
\end{figure}

    Because the mean trapping time is finite for $\alpha>1$, we consider two typical renewal processes; ordinary 
    renewal process and equilibrium renewal process \cite{Cox}.  Equilibrium renewal process is assumed to
  start $-t_a=-\infty$ (see Fig.~\ref{traj_eq}). 
  The PDF of the first jump length $x$ and the first apparent trapping time (the forward recurrence time) $t$,
  $\psi_0(x, t)$, is given by
  \begin{equation}
    \label{e.joint-eq}
    \psi_0(x, t) = \frac{1}{\mu} \int_t^{\infty} w(t')
    \frac {\delta (x-t'^{\gamma}) + \delta (x + t'^{\gamma})}{2}dt',
  \end{equation}
{\color{black}  where $\mu$ is the mean trapping time.
  See Appendix~A} for the derivation. Generally, the first (true) trapping time is longer than the first apparent trapping time. 
  Integrating  Eq.~(\ref{e.joint-eq}) in terms of $x$, we have the PDF of the first apparent 
  trapping time:
  \begin{equation}
    \label{e.trap-time-eq}
    w_0(t) = \frac {1}{\mu} \int_t^{\infty} w(t')dt'.
  \end{equation} 
   Note that the
  Eq.~(\ref{e.trap-time-eq}) is consistent with the result obtained in
  renewal theory, i.e., the PDF of the forward recurrence time \cite{Cox}. Thus, the joint PDF of the first jump length
  and the first apparent trapping time, $\psi_0 (x, t)$, is not given by the form in
  Eq.~(\ref{e.joint-prob}). This is because the first jump length $x$ is
  determined by the time elapsed since a random walker's last jump (the first true trapping time) 
  and thus it is not directly related to the first apparent trapping time, 
  i.e., the time elapsed since the beginning of the measurement at
  $t=0$.  For ordinary renewal process, we just set
  $w_0(t)=w(t)$ and $\psi_0(x,t)=\psi(x,t)$. For $\alpha \leq 1$, we only consider an ordinary renewal process 
  because there is no equilibrium ensemble due to divergent mean trapping time which causes aging \cite{Barkai2003,Schulz2013,Akimoto2013c}.

\section {Generalized Renewal Equation}

 The spacial distribution $P(x,t)$ of CTRWs with starting from the origin
 satisfies the generized renewal equations:
 \begin{align}
  \label{e.recursion}
  P(x,t) &= \int_{0}^{t} dt' \Psi (t-t') Q(x,t') + \Psi_0(t) \delta(x),
  \\[0.1cm]
  Q (x,t) &=
  \int_{-\infty}^{\infty} dx' \int_{0}^{t} dt' \psi (x', t') Q (x-x', t-t')
  \nonumber\\[0.0cm]
  &\quad+ \psi_0 (x, t),
  \end{align}
  where $Q(x,t) dt dx$ is the probability of a random walker reaching an
  interval $[x,x+dx)$ just in a period $[t, t+dt)$, and $\Psi(t)$ [$\Psi_0 (t)$] is the probability of being trapped for longer than
  time $t$ just after a renewal (after the measurement starts at
  $t=0$). $\Psi (t)$ and $\Psi_0 (t)$ are defined as follows:
  \begin{align}
    \label{e.psi-cctrw}
    \Psi (t)
    &=
    1 - \int_{-\infty}^{\infty}dx' \int_{0}^{t} \psi (x', t') dt'
    =
    1 - \int_{0}^{t} w (t') dt',\\[0.2cm]
    \Psi_0 (t)
    &=
    1 - \int_{-\infty}^{\infty}dx' \int_{0}^{t} \psi_0 (x', t') dt'
    =
    1 - \int_{0}^{t} w_0 (t') dt'.
  \end{align}
  Then, the Laplace transforms of these functions are given by
  \begin{equation}
    \label{e.psi-cctrw-laplace}
    \hat{\Psi}(s) = \frac {1 - \hat{w} (s)}{s}, \qquad 
    \hat{\Psi}_0(s) = \frac{\mu s -1 +\hat{w}(s)}{\mu s^2}.
  \end{equation}
  In an ordinary renewal process, $\psi_0(x,t)$ is the same as $\psi (x,t)$. On the other hand, 
  the first jump length $x$ is not determined by the first apparent trapping time $t$ in an equilibrium renewal process, while 
  the first jump length is not independent of the first apparent trapping time.  
  Fourier-Laplace transform with respect to space and time ($x\rightarrow k$
  and $t\rightarrow s$, respectively),  defined by
  \begin{equation} 
  {\hat{P}}(k,s) \equiv \int_{-\infty}^\infty dx \int_0^\infty dt P(x,t) e^{ikx} e^{-st} , 
  \end{equation}
 gives 
  \begin{eqnarray}
    {\hat{P}}(k,s) 
    &=& \frac{\hat{\Psi} (s)\hat{\psi}_0 (k,s) + \hat{\Psi}_0 (s) [1-\hat{\psi} (k,s)]}{1-{\hat{\psi}}(k,s)},
    \label{masterl_eq}
  \end{eqnarray}
  {\color{black}where $\hat{\psi}_0 (k,s)$ and $\hat{\psi} (k,s)$ are Fourier-Laplace transforms of 
  ${\psi}_0 (x,t)$ and ${\psi} (x,t)$, respectively.}
  In what follows, we use the notations $P^{o}(x,t)$ and
  $\hat{P}^{o} (k,s)$ for the ordinary renewal process, and $P^{eq}(x,t)$ and
  $\hat{P}^{eq} (k,s)$ for the equilibrium renewal process. 
  
  For ordinary renewal process, i.e., $\psi_0(t)=\psi(t)$ and
  $\Psi_0(t)=\Psi(t)$, we have the following generalized renewal equation in
  the Fourier and Laplace space:
  \begin{equation}
    \label{e.renewal-cctrw}
    {\hat{P}}^o(k,s) =
    \frac {1}{s}
    \frac{1-\hat{w}(s)}{1-{\hat{\psi}}(k,s)},
  \end{equation}
  where we used Eq.~(\ref{e.psi-cctrw-laplace}).
  For equilibrium renewal process ($\alpha >1$), we have
  \begin{align}
    \label{g.renewal.eq_eq}
    {\hat{P}}^{eq}(k,s) &=
    \frac {1}{s} - \frac {1-\hat{w}(s)}{\mu s^2}
    \frac{1 - \hat{\psi}(k,0)}{1-{\hat{\psi}}(k,s)} ,
  \end{align}
  where we used Eq.~(\ref{e.psi-cctrw-laplace}) and 
  \begin{equation}
    \hat{\psi}_0 (k,s) = \frac{\hat{\psi}(k,0) - \hat{\psi}(k,s)}{\mu s}.
    \label{Laplace_joint_pdf_eq}
  \end{equation}
  The derivation of Eq.~(\ref{Laplace_joint_pdf_eq}) is shown in Appendix~A.
  
  Thus we expressed $\hat{P}^{o}(k,s)$ and $\hat{P}^{eq}(k,s)$
  with $\hat{\psi} (k, t)$ and $\hat{w}(s)$ [Eqs.~(\ref{e.renewal-cctrw})
  and (\ref{g.renewal.eq_eq})]. Now, we derive the explicit forms of these
  functions.
  From Eq.~(\ref{e.joint-prob}), ${\hat{\psi}} (k,s)$ is given by 
  \begin{equation}
  \label{e.joint-prob-fourier-laplace}
  {\hat{\psi}} (k,s) =
  \int_{0}^{\infty} e^{-st} \cos \left( k t^{\gamma}\right) w(t) dt.
  \end{equation}
  Note that ${\hat{\psi}} (0,s) = \hat{w} (s)$. In addition, from
  Eq.~(\ref{e.pdf-trap-jump}), the asymptotic behavior of the Laplace
  transform $\hat{w}(s)$ for $s\rightarrow 0$ is given by 
  \begin{align*}
    \label{e.trap-pdf-laplace}
    \hat{w}(s) =
    \begin{cases}
      1 -  c s^{\alpha} +o(s^\alpha), \quad        & (0<\alpha<1)\\[.3cm]
      1 + c_0 s \ln s +o(s \ln s), \quad  & (\alpha=1) \\[.3cm]
      1 -  \mu s + cs^{\alpha} +o(s^\alpha), \quad & (1<\alpha<2)
    \end{cases}
  \end{align*}
  where $\mu = \left\langle t \right\rangle = \int_0^{\infty} t w(t) dt$.

\section {Mean Square Displacement}

 The asymptotic behavior of the moments of position $x_t$
 for $t\rightarrow \infty$ can be obtained using the Fourier-Laplace transform
 $\hat{P}(k,s)$. Because $\left. \frac{\partial {\hat{P}} (k,s)}{\partial k}\right|_{k=0} =0$, 
 $\langle x_t \rangle=0 $ for both renewal processes. 
  
  In ordinary renewal process, the Laplace transform of the second moment, i.e., 
  the ensemble-averaged MSD (EAMSD), is given by
  \begin{eqnarray}
  \langle x^2_s \rangle_o &=&
  \left. -\frac{\partial^2 {\hat{P}}^o(k,s)}{\partial k^2} \right|_{k=0}
  = -\frac {1}{s} \frac{{\hat{\psi}}''(0,s)}{1 - {\hat{w}}(s)},
  \label{e.eamsd.laplace}
  \end{eqnarray}
  where the ensemble average $\langle \ldots \rangle_o$ is taken with respect to an ordinary 
  renewal process. For $\alpha \in (0,1)$, we obtain the EAMSD \cite{Akimoto2013a}:
  \begin{equation}
  \langle x^2_t \rangle_o \simeq
  \begin{cases}
  \dfrac {\left\langle l^2 \right\rangle}{c \Gamma (1 + \alpha)} t^{\alpha},
    &\quad \left(0 < 2\gamma < \alpha\right)\\[0.5cm]
    \dfrac {1}{|\Gamma(-\alpha)| \Gamma (1 + \alpha)} t^{\alpha} \log t,  
    &\quad \left(2\gamma = \alpha\right)\\[0.5cm]
    \dfrac{\Gamma(2\gamma-\alpha)}{|\Gamma (-\alpha)| \Gamma(1+2\gamma)}  t^{2\gamma}, 
    &\quad  \left(\alpha < 2\gamma \leq 2\right)
  \end{cases}
  \label{eamsd}
  \end{equation}
  where we used $\left\langle t^{2\gamma} \right\rangle = \left\langle l^{2}
  \right\rangle = \int_{-\infty} ^\infty x^2 l(x) dx$ when $\langle l^2 \rangle<\infty$. 

 For $\alpha = 1$, the EAMSD is given by
  \begin{equation}
    \langle x^2_t \rangle_o \simeq
    \begin{cases}
    \dfrac {\left\langle l^2 \right\rangle}{c_0} \dfrac {t}{ \log t},
      &\quad \left(0 < 2\gamma < 1\right) \\[0.5cm]
      t,
      &\quad \left(2\gamma = 1\right)\\[0.3cm]
      \dfrac{\Gamma(2\gamma-1)}{\Gamma(2\gamma +1)} \dfrac {t^{2\gamma}}{ \log t},
      &\quad \left(1 < 2\gamma \leq  2\right)
    \end{cases}
    \label{eamsd2}
  \end{equation}

 Finally, for $\alpha \in (1,2)$, the EAMSD  is given by
 \begin{equation}
    \langle x^2_t \rangle_o \simeq
    \begin{cases}
      \frac {\left\langle l^2 \right\rangle}{\mu} t \left[1 +\frac{ct^{1-\alpha}}{\mu \Gamma(3-\alpha)} \right],
      & \left(0 < 2\gamma <  1\right) \\[0.5cm]
      \frac {\left\langle l^2 \right\rangle}{\mu} t +
      \frac {ct^{2-\alpha}}{\mu \Gamma (3 - \alpha)}
      \left[
      \frac {\left\langle l^2 \right\rangle}{\mu} -\alpha
      \right],
      &  \left(2\gamma =  1\right) \\[0.5cm]
      \frac {\left\langle l^2 \right\rangle}{\mu} t 
      + \frac {c_0}{\mu} \frac {\Gamma (2\gamma - \alpha)}{\Gamma (2\gamma - \alpha + 2)} t^{2\gamma-\alpha +1},
      & \left(1 < 2\gamma <  \alpha\right) \\[0.5cm]
      \frac {c_0}{\mu} t \log t,
      & \left(2\gamma = \alpha\right) \\[0.5cm]
      \frac{c_0\Gamma(2\gamma-\alpha)}{\mu \Gamma(2\gamma-\alpha +2)}t^{2\gamma-\alpha +1}.
      & \left(\alpha < 2\gamma \leq  2\right)
    \end{cases}
    \label{eamsd3}
  \end{equation}
  {\color{black} These results are consistent with a previous study \cite{Klafter1987}.
  We note that the EAMSD for $\gamma=1$ is smaller than that in L\'{e}vy walk, whereas the 
  scaling exponent $3-\alpha$ is the same as that in L\'{e}vy walk \cite{Zumofen1993}. This is because 
  SEDLF is a wait and jump model, while L\'{e}vy walk is a moving model.}
  
  In equilibrium renewal process ($\alpha >1$),
  \begin{align}
    \label{e.eamsd-eq-laplace}
    \langle x^2_s \rangle_{eq} &=
    \left. -\frac{\partial^2 {\hat{P}}^{\rm eq}(k,s)}{\partial k^2} \right|_{k=0}
    = \frac {\hat{\psi}''(0,0)}{\mu s^2}.
  \end{align}
  Eq.~(\ref{e.eamsd-eq-laplace}) is valid only for $0 < 2\gamma < \alpha-1$, 
  otherwise the second moment of the first jump length diverges, i.e., the EAMSD diverges. {\color{black}This is very different from 
  L\'{e}vy walk process because there exists an equilibrium renewal process in L\'{e}vy walk with $\alpha >1$. }
  Because $\hat{\psi}''(0,0) = \left\langle l^2 \right\rangle$ for $0 < 2\gamma < \alpha-1$, the EAMSD is given by
  \begin{equation}
  \langle x^2_t \rangle_{eq} = \frac{\langle l^2 \rangle}{\mu} t.
  \label{eamsd-eq}
  \end{equation}
  {\color{black}  In SEDLF, the leading order of the EAMSD in an ordinary renewal process is the same as that in an equilibrium renewal process 
  ($\alpha > 2\gamma +1$). On the other hand, in L\'{e}vy walk, the proportional constant of the EAMSD 
  in a non-equilibrium ensemble such as an ordinary renewal process 
  differs from that in an equilibrium one \cite{Froemberg2013,Zumofen1993,Godec2013,Froemberg2013b}. We note that the TAMSD coincides with
  the EAMSD in an equilibrium ensemble as the measurement time goes to infinity. 
  Significant initial ensemble dependence of statistical quantity has been also observed in non-hyperbolic dynamical systems \cite{Akimoto2007}.}

\section {Time-averaged Mean Square Displacement}
In normal Brownian motion, the TAMSD defined by Eq.~(\ref{tamsd}) converges to 
the MSD {\color{black}with an equilibrium ensemble}:
\begin{equation}
\overline{\delta^2 (\Delta; t)} \rightarrow {\color{black}\langle x_{\Delta} ^2 \rangle_{eq}}~{\rm as}~t\to \infty. 
\end{equation}
Such ergodic property does not hold in various stochastic models of anomalous diffusion such as 
CTRW and L\'evy walk \cite{Lubelski2008,He2008,Froemberg2013}.
Here, we derive the TAMSD in the SEDLF. 
In {\color{black}wait and jump random walks with random waiting times} such as CTRW and SEDLF, the TAMSD can be represented using the total number of
jumps, denoted by $N_t$, and $h_k = \Delta
l_k^2 + 2\sum_{m=1}^{k-1}l_kl_m \theta (\Delta - t_k+t_m)$ \cite{Miyaguchi2011a,Akimoto2013a,Miyaguchi2013}:
\begin{equation}
  \overline{\delta^2 (\Delta; t)} \simeq \frac{1}{t}\sum_{k=0}^{N_t} h_k\quad (t\rightarrow\infty),
  \label{tamsd_Nt}
\end{equation}
where $l_k$ is the $k$-th jump length, $t_k$ is the time when
  the $k$-th jump occurs,
and $\theta(t)$ is a step function, defined by $\theta(t)=0$ for $t<0$ and
$t$ otherwise.
For $\gamma>0$, one can show that 
$\sum_{k=0}^{N_t} (h_k - \Delta l^2_k)/\sum_{k=0}^{N_t} l_k^2 \rightarrow 0$ as $t\rightarrow \infty$ \cite{Akimoto2013a}. 
Therefore, the TAMSD can be written as 
\begin{equation}
  \label{tamsd_Nt2}
  \overline{\delta^2 (\Delta; t)} \simeq D_t \Delta 
  \quad (\Delta \ll t ~{\rm and}~t\rightarrow \infty),
\end{equation}
where $D_t = \sum_{k=0}^{N_t} l^2_k/t$. {\color{black}We note that the relation (\ref{tamsd_Nt}) does not hold if the random 
walker moves with constant speed as in L\'{e}vy walk because $(x_{t+\Delta} - x_t)^2$ is simply zero in SEDLF but not 
in L\'{e}vy walk. In fact, the TAMSD does not increase linearly with time in L\'{e}vy walk \cite{Froemberg2013,Godec2013,Froemberg2013b}.}

To investigate an ergodic property of the time-averaged diffusion coefficient $D_t$, we derive the PDF $P_2 (z,t)$ of $Z_t\equiv
\sum_{k=0}^{N_t} l^2_k$.
Because $l^2_k$  and $N_t$ are mutually correlated, we use the generalized renewal equation  
 for $Z_t$: 
 \begin{align}
  \label{e.recursion.z.1}
  P_2(z,t) &= \int_{0}^{t} dt' \Psi (t-t') Q_2(z,t') + \Psi_0(t) \delta(z),
  \\[0.1cm]
  \label{e.recursion.z.2}
  Q_2 (z,t) &=
  \int_{0}^{\infty} dz' \int_{0}^{t} dt' \phi (z', t') Q_2 (z-x', t-t')
  \nonumber\\[0.0cm]
  &\quad
  + \phi_0 (z, t),
\end{align}
 where the joint PDF $\phi(z,t)$ is given by 
\begin{equation}
  \label{e.joint-prob-z}
  \phi (z,t) = w(t) \delta(z-t^{2\gamma})
\end{equation}
The joint PDF of the first squared jump length $z=l_0^2$ and apparent
  trapping time $t$, $\phi_0 (z,t)$, is given by $\phi_0 (z,t) = \phi (z,t)$
  for the ordinary ensemble, whereas
  \begin{equation}
    \phi_0 (z,t) = \frac {1}{\mu} \int_t^{\infty} dt' w(t') \delta(z-t'^{2\gamma}),
  \end{equation}
  for equilibrium ensemble, which can be derived in the same way as
  the derivation of $\psi_0 (x,t)$ given in Appendix~A. The double Laplace
  transform with respect to $z$ and time $t$ is defined by
  \begin{equation} 
    {\hat{P}}_2(k,s) \equiv \int_{0}^\infty dz \int_0^\infty dt P_2(z,t) e^{-kz} e^{-st}.
    \label{masterl_eq_z}
  \end{equation}
  From the generalized renewal equations (\ref{e.recursion.z.1}) and
  (\ref{e.recursion.z.2}), we obtain 
  \begin{equation} 
    {\hat{P}}_2(k,s) =
    \frac{\hat{\Psi} (s)\hat{\phi}_0 (k,s) + \hat{\Psi}_0 (s) [1-\hat{\phi} (k,s)]}{1-{\hat{\phi}}(k,s)},
    \label{masterl_eq_z}
  \end{equation}
  where the double Laplace transform of $\phi (z, t)$ is given by
  \begin{equation}
    \hat{\phi} (k,s) = \int_{0}^{\infty} e^{-st} e^{-kt^{2\gamma}} w(t) dt,
  \end{equation}
  and that of $\phi_0 (z,t)$ is given by $\phi_0 (z, t) = \phi (z,t)$ for
  the ordinary process, and by 
  \begin{align}
    \hat{\phi}_{0} (k,s) &=
    \frac {\hat{\phi}(k,0) - \hat{\phi} (k, s)}{\mu s},
  \end{align}
  for the equilibrium process.  Note that $\hat{\phi} (0, s) = \hat{w}(s)$.

\subsection{Ordinary Renewal Process}

In ordinary renewal process,  the Laplace transform ${\hat{P}_2}^o(k,s)$ is given by
\begin{equation}
  \label{e.renewal-cctrw-z}
  {\hat{P}_2}^o(k,s) =
  \frac {1}{s}
  \frac{1-\hat{w}(s)}{1-{\hat{\phi}}(k,s)}.
\end{equation}
Thus, we have the Laplace transform of $\langle Z_t \rangle_o$
  as follows:
  \begin{equation}
    \langle Z_s \rangle_o
    = \left.-\frac{\partial \hat{P}^o_2(k,s)}{\partial k} \right|_{k=0}
    = \frac{- \hat{\phi}'(0,s)}{s [ 1- \hat{w}(s)]}
    = \left\langle x_s^2 \right\rangle_o,
  \end{equation}
  where we used $- \hat{\phi}'(0,s) = - \hat{\psi}''(0,s)$ and
  Eq.~(\ref{e.eamsd.laplace}). Then, averaging Eq.~(\ref{tamsd_Nt2}) over an 
  ordinary ensemble, we have
  \begin{equation}
    \langle \overline{\delta^2 (\Delta; t)} \rangle_o \simeq
    \left\langle D_t \right\rangle_o \Delta=
    \frac{\langle x_t^2 \rangle_o}{t} \Delta 
  \end{equation}
  Therefore, using Eqs.~(\ref{eamsd})--(\ref{eamsd3}), we have the leading
  terms of the mean diffusion coefficient, $\langle D_t \rangle_o = \langle
  x_t^2 \rangle_o/t$, for $t\rightarrow \infty$ as follows:
  \begin{align}
    \langle D_t \rangle_o
    &\simeq 
    \begin{cases}
      \frac{\left\langle l^2 \right\rangle }{c \Gamma(1+\alpha)} t^{\alpha-1}
      & \left(0 < 2\gamma <  \alpha\right)\\[0.3cm]
      \frac{1}{|\Gamma (-\alpha)| \Gamma(1+\alpha)} t^{\alpha-1} \log t
      & \left(2\gamma = \alpha\right) \\[0.3cm]
      \frac{\Gamma(2\gamma-\alpha)}{|\Gamma (-\alpha)| \Gamma(1+2\gamma)} t^{2\gamma-1}
      & \left(\alpha < 2\gamma \leq 2\right) 
    \end{cases}
    \label{e.mean_diffusion1}
  \end{align}
  for $0 < \alpha < 1$,
  \begin{equation}
    \langle D_t \rangle_o \simeq
    \begin{cases}
      \frac {\left\langle l^2 \right\rangle}{c_0} \frac {1}{ \log t} +o\left(\frac{1}{\log t}\right).
      &\quad \left(0 < 2\gamma < 1\right)\\[0.3cm]
      1 + O\left(\frac{1}{\log t}\right),
      &\quad \left(2\gamma = 1\right)\\[0.3cm]
      \frac{\Gamma(2\gamma-1)}{\Gamma(2\gamma +1)} \frac {t^{2\gamma-1}}{ \log t},
      &\quad \left(1 < 2\gamma \leq 2\right)
    \end{cases}
    \label{e.mean_diffusion2}
  \end{equation}
  for $\alpha = 1$, and
  \begin{equation}
    \langle D_t \rangle_o \simeq
    \begin{cases}
      \frac {\left\langle l^2 \right\rangle}{\mu}\left[1 +\frac{ct^{1-\alpha}}{\mu \Gamma(3-\alpha)} \right],
      & \left(0 < 2\gamma < 1\right) \\[0.3cm]
      \frac {\left\langle l^2 \right\rangle}{\mu} +
      \frac {ct^{1-\alpha}}{\mu \Gamma (3 - \alpha)}
      \left[
      \frac {\left\langle l^2 \right\rangle}{\mu} -\alpha
      \right],
      & \left(2\gamma = 1\right) \\[0.3cm]
      \frac {\left\langle l^2 \right\rangle}{\mu} 
      + \frac {c_0}{\mu} \frac {\Gamma (2\gamma - \alpha)}{\Gamma (2\gamma - \alpha + 2)} t^{2\gamma-\alpha},
      & \left(1 < 2\gamma < \alpha\right)\\[0.3cm]
      \frac {c_0}{\mu} \log t,
      & \left(2\gamma = \alpha\right)\\[0.3cm]
      \frac{c_0\Gamma(2\gamma-\alpha)}{\mu \Gamma(2\gamma-\alpha +2)}t^{2\gamma-\alpha}
      & \left(\alpha < 2\gamma \leq 2\right)
    \end{cases}
    \label{e.mean_diffusion3}
  \end{equation}
  for $1 < \alpha < 2$. It follows that the mean diffusion coefficient
  diverges as $t$ goes to infinity for $1<\alpha<2$ and $\alpha< 2\gamma < 2$.

 Similarly, the Laplace transform of $\langle Z_t^2 \rangle$ is given by 
 \begin{equation}
 \langle Z_s^2 \rangle_o = \frac{1}{s [ 1- \hat{w}(s)]} \left\{ \frac{2\hat{\phi}'(0,s)^2}{1-\hat{w}(s)} + \hat{\phi}''(0,s)\right\}.
 \end{equation}
 It follows that the leading order of  the second moment of $D_t$ is given by
 \begin{equation}
  \langle D^{2}_t \rangle_o \simeq
  \begin{cases}
    \frac {2 \left\langle l^2 \right\rangle^2}{c^2 \Gamma (2 \alpha +1)} t^{2(\alpha-1)},
    & (0< 2 \gamma < \alpha)\\[0.2cm]
    \frac {2}{\Gamma(2\alpha + 1) |\Gamma (-\alpha)|} \left(t^{\alpha} \log t\right)^{2},
    & (2 \gamma = \alpha) \\[0.2cm]
    \frac{\Gamma(4\gamma-\alpha) |\Gamma (-\alpha)| + 2\Gamma(2\gamma-\alpha)^2}{\Gamma(4\gamma+1)|\Gamma(-\alpha)|^2}
    t^{4\gamma-2},
    & (\alpha < 2 \gamma \leq 2) 
  \end{cases}
  \end{equation}
  for $0 < \alpha <1$, 
  \begin{equation}
    \langle D_t^2 \rangle_o \simeq
    \begin{cases}
      \frac {\left\langle l^2 \right\rangle^2}{c_0^2} \frac {1}{(\log t)^2} + o\left(\frac {1}{(\log t)^2}\right),
      &\quad \left(0 < 2 \gamma < 1\right)\\[0.3cm]
      1  + O\left(\frac {1}{\log t}\right),
      &\quad (2\gamma =  1)\\[0.3cm]
      \frac{\Gamma(4\gamma-1)}{\Gamma(4\gamma +1)} \frac {t^{4\gamma-2}}{ \log t}.
      &\quad \left(1 < 2\gamma \leq 2\right)
    \end{cases}
    \label{e.mean_diffusion2}
  \end{equation}
  for $\alpha = 1$, and 
  \begin{equation}
    \langle D^{2}_t \rangle_o \simeq
    \begin{cases}
      \frac {\left\langle l^2 \right\rangle^2}{\mu^2}
      + \frac {4c\left\langle l^2 \right\rangle^2}{\mu^3}
      \frac {t^{1-\alpha}}{\Gamma(4 - \alpha)},
      & \left(0 < 2\gamma < 1\right)\\[0.2cm] 
      \frac {\left\langle l^2 \right\rangle^2}{\mu^2}
      + \frac {Kt^{1-\alpha}}{\Gamma(4 - \alpha)},
      & \left(2\gamma = 1\right) \\[0.2cm]
      \frac {\left\langle l^2 \right\rangle^2}{\mu^2}
      + \frac {c_0}{\mu}
      \frac {\Gamma (4\gamma - \alpha)}{\Gamma(4\gamma - \alpha + 2)}
      t^{4\gamma-\alpha-1},
      & \left(1 < 2\gamma < \frac {\alpha+1}{2}\right)\\[0.2cm] 
      \frac {\left\langle l^2 \right\rangle^2}{\mu^2}
      + \frac {c_0}{2\mu},
      & (2\gamma = \frac {\alpha+1}{2})\\[0.2cm] 
      \frac{c_0 \Gamma(4\gamma-\alpha)  }{\mu \Gamma(4\gamma -\alpha +2)}
      t^{4\gamma-\alpha -1},
      & \left(\frac{1+\alpha}{2} < 2\gamma \leq 2\right)  \\[0.2cm]
    \end{cases}
    \label{2nd_diffusion}
  \end{equation}  
  for $\alpha >1$ $\left(K := \frac {4\left\langle l^2 \right\rangle c}{\mu^2}
  \left\{\frac {\left\langle l^2 \right\rangle}{\mu} - \alpha\right\} +
  \frac {c_0 \Gamma (2 - \alpha)}{\mu}\right)$.
  
\begin{figure}
  \includegraphics[width=.9\linewidth, angle=0]{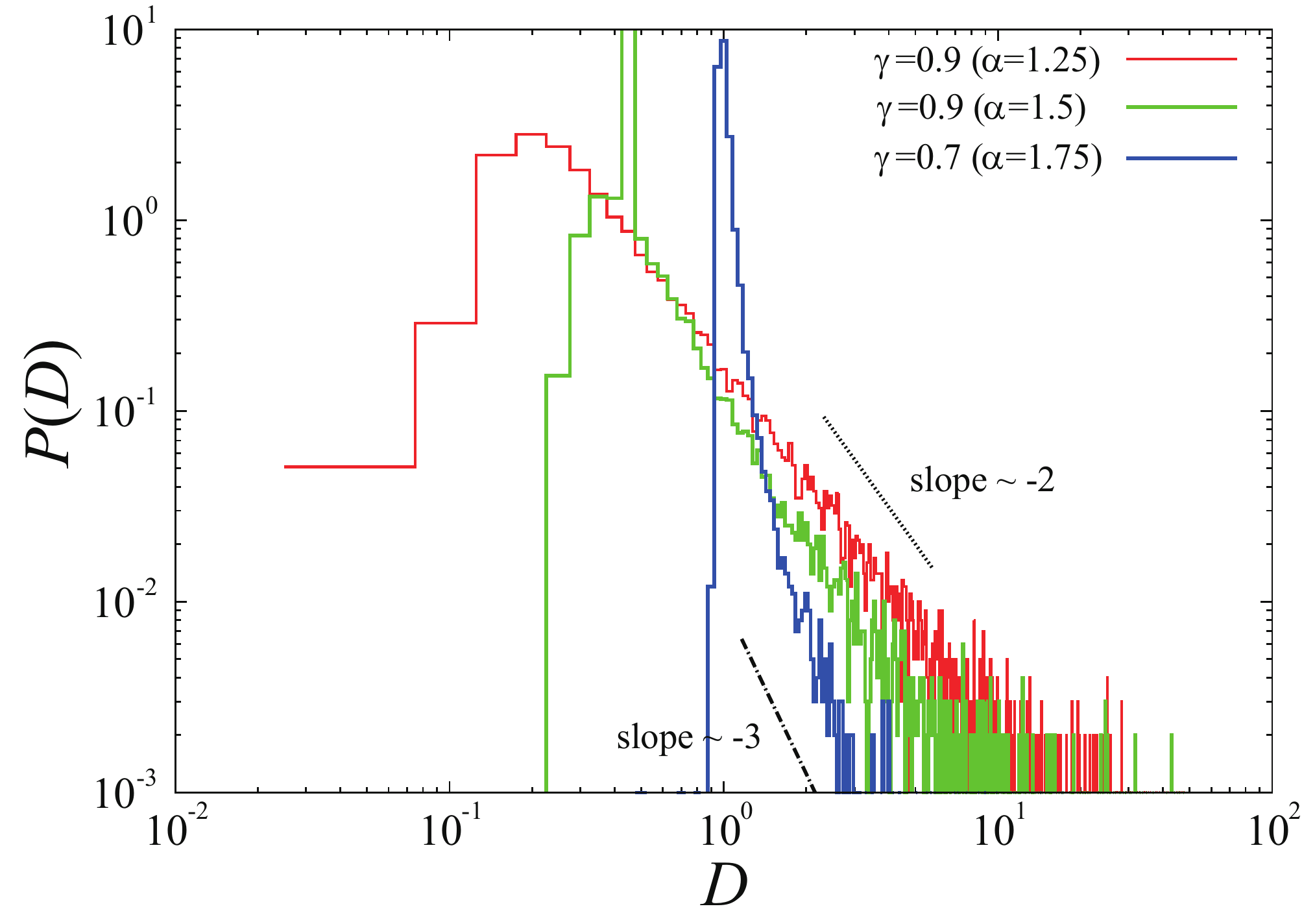}
  \caption{ Histograms of the normalized diffusion coefficients $D\equiv D_t/\langle D_t \rangle$  for different $\gamma$ and $\alpha$ ($t=10^6$). 
  We calculate $D_t$ by $\overline{\delta^2 (\Delta; t)}/\Delta$ in numerical simulations with $\Delta = 10$. 
  The dashed line segments represent power-law distributions with exponent  $-2$ and $-3$ for reference. 
  }
  \label{pdfd}
\end{figure}

  Now, we study the relative standard deviation (RSD) of $D_t$,
  $\sigma_{\rm EB} (t) \equiv \sqrt{\langle D_t^2 \rangle -
    \langle D_t \rangle^2}/\langle D_t \rangle$ \cite{He2008,Miyaguchi2011,Miyaguchi2011a}, to measure
  the ergodicity breaking.  First, for $0< \alpha <1$, RSD
  $\sigma_{\mathrm{EB}} (t)$ does not converge to zero but to a finite value as
  $t\rightarrow \infty$.  Therefore, the diffusion coefficients remain
  random even when the measurement time goes to infinity
  \cite{Akimoto2013a}.  Second,
  for $\alpha = 1$, we expect usual ergodic
  behavior for $0 < 2\gamma \leq 1$, because the RSD goes to zero as $t\to \infty$.
   However, for $1<2 \gamma <2$,
  the RSD diverges as $t \to \infty$:
  \begin{equation*}
    \sigma_{\mathrm{EB}}^2 (t) \sim
    \log t.
  \end{equation*}
  Finally, for $1 < \alpha <2$, we have
  \begin{align}
    \sigma_{\mathrm{EB}}^2 (t) \sim
    \begin{cases}
      t^{1-\alpha}, \qquad
      & \left(0 < 2\gamma \leq 1\right)\\[.2cm]
      t^{4\gamma -\alpha - 1}, \qquad
      &\left(1 < 2\gamma < \frac {\alpha+1}{2}\right)\\[.2cm]
      \frac {\mu c_0}{2\left\langle l^2 \right\rangle^2}, \qquad
      &\left(2\gamma = \frac {\alpha+1}{2}\right)\\[.2cm]
      t^{4\gamma -\alpha - 1}, \qquad
      &\left(\frac {\alpha+1}{2} < 2\gamma <  \alpha\right)\\[.2cm]
      \frac {t^{\alpha - 1}}{(\log t)^{2}}, \qquad
      &(2\gamma =  \alpha)\\[.2cm]
      t^{\alpha - 1}, \qquad
      &\left(\alpha < 2\gamma \leq 2\right)
    \end{cases}
  \end{align}
  Thus, {\color{black} TAMSDs show ergodic behavior when the parameters satisfy $0 < 2 \gamma < \frac {\alpha+1}{2}$ because 
  the RSD goes to zero as $t\to \infty$. On the other hand,} the RSD  
  converges to a finite value
  for $\gamma = \frac{\alpha+1}{4}$, and diverges for $\frac {\alpha+1}{2} < 2 \gamma \leq 2$.  
  Numerical simulations suggest that this divergence of the RSD for the case of $\frac{\alpha+1}{2} < 2 \gamma \leq 2$ 
  will be attributed to {\color{black}a power-law with divergent second moment in the PDF of $D_t/\langle D_t \rangle_o$ 
   (see Fig.~\ref{pdfd}).  Because the RSD is defined using the second moment of $D_t$, it diverges, 
    whereas the PDF converges to a power-law distribution.}
   Thus, the RSD, 
  $\sigma_{\mathrm{EB}} (t)$, is not helpful to characterize the ergodicity
  breaking in this case. Using the relative fluctuation defined by $R(t) \equiv  \langle |D_t -
    \langle D_t \rangle |\rangle /\langle D_t \rangle$ \cite{Akimoto2011,Uneyama2012} instead of the RSD, we can clearly see the ergodicity breaking 
    in the case of $\frac{\alpha+1}{2} < 2 \gamma \leq 2$: we numerically found that the relative fluctuation, 
    {\color{black}$R(t) =  \langle |D_t/\langle D_t \rangle_o - 1 |\rangle_o$,} 
    converges to a constant as $t\to \infty$ because {\color{black}$D_t/\langle D_t \rangle_o$} does not converge to one 
    but converges in distribution. Thus, the ergodicity {\color{black}in TAMSD}
    breaks down for $\frac{\alpha+1}{2} \leq 2 \gamma \leq2$.

 For $\alpha >1$, the asymptotic behavior of the Laplace transform of $\langle Z_t^n \rangle_o$ at $s\to 0$ ($n>1$) is given by 
 \begin{equation}
 \langle Z_s^n \rangle_o \simeq 
 \begin{cases}
 (-1)^n\dfrac{\hat{\psi}_2^{(n)}(0,s)}{s [ 1- \hat{w}(s)]},
 & (2\gamma \geq \frac{1+\alpha}{2}) \\[0.4cm] 
 (-1)^n\dfrac{n! \langle l^2 \rangle^n}{\mu^n}.
 & (2\gamma < \frac{1+\alpha}{2})
 \end{cases}
 \end{equation}
 It follows that the leading order of  the $n$th moment of $D_t$ is given by
 \begin{equation}
 \langle D_t^n \rangle_o \simeq 
 \begin{cases}
 \dfrac{c_0 \Gamma(4n\gamma-\alpha)  }{\mu \Gamma(4n\gamma -\alpha +2)}
    t^{2n\gamma-\alpha -1},
    & (2\gamma \geq \frac{1+\alpha}{2}) \\[0.4cm] 
    (-1)^n\dfrac{n! \langle l^2 \rangle^n}{\mu^n}.
    & (2\gamma < \frac{1+\alpha}{2})
 \end{cases}
 \end{equation}
 By numerical simulations, we confirm that  
 the scaled diffusion coefficient {\color{black}$D_t/\langle D_t \rangle_o$} converges in distribution to a random variable $S_{\alpha,\gamma}$:
 \begin{equation}
 \frac{D_t}{\color{black}\langle D_t \rangle_o}\Rightarrow S_{\alpha,\gamma} ~{\rm as}~t\to \infty,
 \end{equation} 
 for $\frac{\alpha + 1}{2} < 2\gamma \leq2$. The distribution depends on $\gamma$ and $\alpha$ (see Fig.~\ref{pdfd}).
 {\color{black}We note that the distribution obeys a power-law with divergent second moment because  
all the $n$-th  moments $(n>1)$ of $D_t/\langle D_t \rangle_o$ diverge as $t$ goes to infinity. In fact, 
as shown in Fig.~2, the power-law exponents are smaller than 3.}

For $\alpha \leq 1$, we obtained all the higher
moments of $D_t$ \cite{Akimoto2013a}. In particular, for $2\gamma < \alpha$, all the moments are given by 
\begin{equation}
{\color{black}\langle D^n_t \rangle_o} \simeq 
\frac{n!\left\langle l^2 \right\rangle^n }{c^n \Gamma(n+\alpha)} t^{n(\alpha-1)}.
\end{equation}
Therefore, the distribution of the scaled diffusion coefficient converges to the Mittag-Leffler distribution:
\begin{equation}
\frac{D_t}{\color{black}\langle D_t \rangle_o} \Rightarrow M_\alpha ~(t\to \infty),
\end{equation} 
where
\begin{equation}
\langle e^{zM_\alpha} \rangle = \sum_{n=0}^\infty \frac{\Gamma(1+\alpha)^nz^n}{\Gamma(1+n\alpha)}.
\end{equation}
Moreover, for $2\gamma > \alpha$, the distribution of $\frac{D_t}{\color{black}\langle D_t \rangle_o}$ also converges to a time-independent non-trivial
distribution as $t\to \infty$ \cite{Akimoto2013a}:
\begin{equation}
\frac{D_t}{\color{black}\langle D_t \rangle_o} \Rightarrow Y_{\alpha,\gamma} ~(t\to \infty).
\end{equation}

\subsection {Equilibrium Renewal Process}

 In an equilibrium renewal process, i.e., $1 < \alpha <2$ and $0< 2\gamma <\alpha-1$, the Laplace
 transform ${\hat{P}_2}^{eq}(k,s)$ is given by
  \begin{equation}
    \label{e.renewal-cctrw-z}
    {\hat{P}_2}^{eq}(k,s) =
    \frac {1}{s}
    -
    \frac{1-\hat{w}(s)}{\mu s} \frac {1-{\hat{\phi}}(k,0)} {1-{\hat{\phi}}(k,s)}. 
  \end{equation}
  Thus, we have the Laplace transform of $\langle Z_t \rangle_{eq}$ as follows:
  \begin{equation}
    \langle Z_s \rangle_{eq}
    = \left.-\frac{\partial \hat{P}^{eq}_2(k,s)}{\partial k} \right|_{k=0}
    = \frac{\left\langle l^2 \right\rangle}{\mu s^2}
    = \left\langle x_s^2 \right\rangle_o.
  \end{equation}
  Averaging Eq.~(\ref{tamsd_Nt2}) over equilibrium ensemble, we have
  \begin{equation}
    \langle \overline{\delta^2 (\Delta; t)} \rangle_{eq} \simeq
    \left\langle D_t \right\rangle_{eq} \Delta=
    \frac{\langle Z_t \rangle_{eq}}{t} \Delta =
    \frac{\left\langle l^2 \right\rangle}{\mu} \Delta, 
  \end{equation}
  where the second moment of the first jump length is finite for $0<\gamma< \alpha-1$, 
  otherwise the ensemble average of the TAMSD diverges. The second moment of the
  diffusion constant is also derived in the similar way:
  \begin{equation}
    \langle Z_s^2 \rangle_{eq}
    = \left.\frac{\partial^2 \hat{P}^{eq}_2(k,s)}{\partial k^2} \right|_{k=0}
    \simeq
    \frac {2\left\langle l^2 \right\rangle^2}{\mu^2 s^3} +
    \frac {2c \left\langle l^2 \right\rangle^2}{\mu^3 s^{4-\alpha}},
  \end{equation}
  and thus we have
  \begin{equation}
    \left\langle D_t^2 \right\rangle_{eq} = 
    \frac {\left\langle l^2 \right\rangle^2}{\mu^2} +
    \frac {2c \left\langle l^2 \right\rangle^2}{\mu^3 \Gamma(4-\alpha)} t^{1-\alpha},
  \end{equation}
  for $0<\gamma< \alpha-1$. From these results, RSD is given by
  \begin{equation}
    \sigma^2_{\mathrm{EB}} (t) \simeq
    \frac {2c}{\mu} \frac {t^{1-\alpha}}{\Gamma (4 - \alpha)}.
  \end{equation}

\begin{figure}
  \includegraphics[width=.9\linewidth, angle=0]{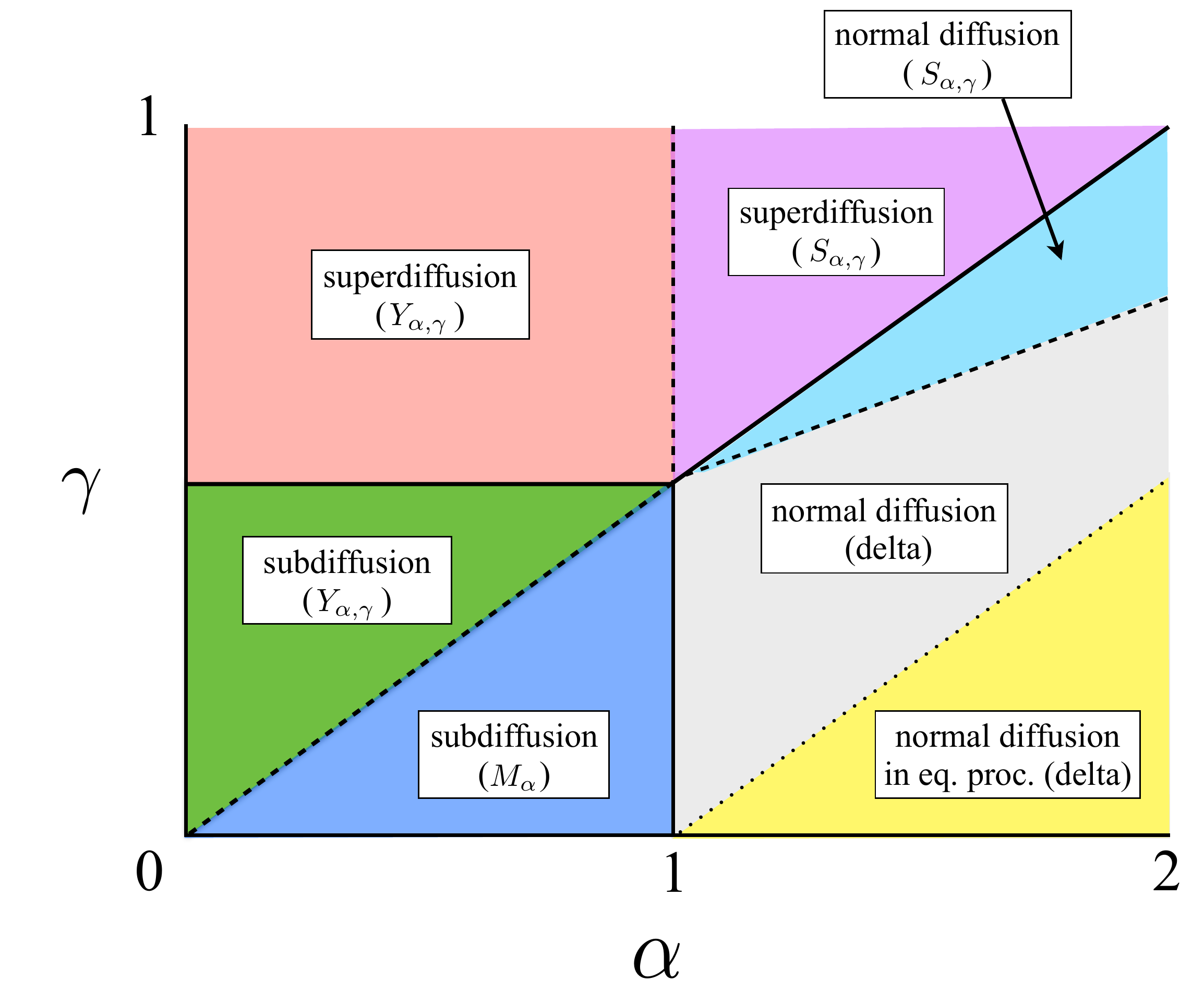}
  \caption{ Phase diagram of the SEDLF in ordinary and equilibrium renewal processes. 
  Solid lines divide the phase of the EAMSD in the ordinary renewal process. The dashed lines divide the phase of the distribution function 
  of the TAMSD. Only below the dotted line $2\gamma < \alpha -1$, the equilibrium process exists.}
  \label{phase}
\end{figure}

{\color{black}
\section{Discussion}

We have shown the phase diagram based on the power-law exponent of anomalous diffusion and the distribution of TAMSDs in SEDLF.
Although SEDLF is closely related to CTRW, L\'{e}vy walk, and L\'{e}vy flight, its statistical properties on anomalous diffusion are different 
form them.  In particular, while the visit points in SEDLF are the same as the turning points of a random walker in L\'{e}vy walk \cite{Shlesinger1987}, 
a random walker cannot move while it is trapped in SEDLF, which is completely different from L\'{e}vy walk. This discrepancy makes 
the scaling of TAMSD different.  
In fact, TAMSDs in SEDLF increase linearly with time even when the EAMSD shows superdiffusion. On the other hand, 
TAMSDs show superlinear scaling (superdiffusion) in L\'{e}vy walk. Because a particle is always moving and there is a power-law coupling between 
waiting times and moving length in turbulent diffusion and diffusion of cold atoms \cite{Shlesinger1987,Barkai2014}, a moving model
of SEDLF can be applied to them. On the other hand, a wait and jump model (SEDLF) will be important 
   in finance  and earthquakes.  
   In particular, SEDLF will be applied to describe dynamics of energy released in earthquake  because 
   energy is gradually accumulated and released in earthquake. Since TAMSDs can not be represented by Eq.~(\ref{tamsd_Nt}) in a moving 
   model such as L\'{e}vy walk and a moving model of SEDLF, 
to investigate ergodic properties of TAMSDs in a moving model of SEDLF is left for future work.
}

\section {Conclusion}

In conclusion, we have shown the phase diagram in SEDLF for a wide range of parameters, where 
the EAMSD shows normal diffusion, subdiffusion and superdiffusion, and the distribution of 
the TAMSD depends on the power-law exponent $\alpha$ of the trapping-time distribution 
as well as the coupling parameter $\gamma$.   
We consider two typical processes: ordinary renewal process and equilibrium renewal process. 
An equilibrium distribution for the first renewal time (the forward recurrence time) 
exists in renewal processes when the mean of interoccurrence time between successive renewals 
does not diverge. However, even when the mean does not diverge, an equilibrium distribution does not exist in the SEDLF
 because of divergence of the second moment of the first jump length. Therefore, 
we have found that the TAMSDs remain random in some parameter region even when the mean trapping time 
does not diverge. In particular, it is interesting to note that this distributional ergodicity is observed even when the EAMSD shows a normal 
diffusion, i.e., $\frac{\alpha+1}{2} < 2\gamma <  \alpha$ and $1<\alpha<2$. 
In this regime, both the mean trapping time and the second moment of jump length are finite. Therefore, this result
  provides a novel route to the distributional ergodicity, because so far
  the distributional ergodicity has been found only in systems with the
  divergent mean trapping time or the divergent second moment of jump
  length, which break down the law of large numbers.

\section*{acknowledgement}
This work was partially supported by Grant-in-Aid for Young Scientists (B) (Grant No. 26800204).

\appendix

\section{Derivation of $\psi_0(x,t)$}
Here, we derive the joint PDF $\psi_{0}(x,t)$ of the first jump length $x$ and the first apparent trapping time $t$ 
for an equilibrium renewal process 
($\alpha >1$). 
Let $\psi_{0,n}(x,t; t_a)$ be the joint PDF that the first jump and the first apparent trapping time after time $t_a$ 
given that the number of jumps in $[0,t_a]$ is $n$. The joint PDF can be 
represented by
\begin{widetext}
\begin{align}
    \psi_{0,n}(x,t; t_a) =&
    \frac{1}{2} \langle \delta (x- (t_{n+1}-t_n)^\gamma) \delta (t - (t_{n+1}- t_a)) I(t_n<t_a<t_{n+1})
    \rangle \\
    &+ 
    \frac{1}{2} \langle \delta (x + (t_{n+1}-t_n)^\gamma) \delta (t - (t_{n+1}- t_a)) I(t_n<t_a<t_{n+1})
    \rangle
  \end{align}
The Fourier and double Laplace transform with respect to $x,t$ and $t_a$ is given by
\begin{eqnarray}
      \hat{\psi}_{0,n}(k,s; s_a) &\equiv&
      \int_{-\infty}^\infty dx \int_0^\infty dt \int_0^\infty dt_a  e^{ikx}e^{-st}e^{-s_at_a} \psi_{0,n}(x,t; t_a)\\
      &=& \left\langle 
      \int_{t_n}^{t_{n+1}} dt_a
      \frac{e^{ik\tau_{n+1}^\gamma} + e^{-ik \tau_{n+1}^\gamma}}{2} e^{-s(t_{n+1}-t_a)}e^{-s_at_a} \right\rangle\\
      &=& \frac{1}{s-s_a}
      \langle e^{-s_a t_n}\rangle \left\langle (e^{-s_a \tau_{n+1}} - e^{-s
        \tau_{n+1}})
      \cos (k\tau_{n+1}^\gamma) \right\rangle\\
      &=& \frac{1}{s-s_a} \hat{w}(s_a)^n [\hat{\psi}(k,s_a) - \hat{\psi}(k,s)].
    \end{eqnarray}
\end{widetext}
Therefore, the Fourier and double Laplace transform of the joint PDF of the first jump length $x$ and 
the first apparent trapping time $t$ after time $t_a$, 
is given by
\begin{eqnarray}
\hat{\psi}_0(k,s;s_a) &=&\sum_{n=0}^\infty \hat{\psi}_{0,n}(k,s; s_a) \\
&=&\frac{1}{s_a-s} \frac{\hat{\psi}(k,s_a) - \hat{\psi}(k,s)}{1-\hat{w}(s_a)}.
\end{eqnarray}
Therefore, the Laplace transform of $\psi_{0}(x,t)$ is given by 
\begin{eqnarray}
\hat{\psi}_0(k,s) &=& \lim_{s_a\to 0} s_a \hat{\psi}_0(k,s;s_a) \\
&=& \frac{\hat{\psi}(k,0) - \hat{\psi}(k,s)}{\mu s}.
\end{eqnarray}
Through the inverse Fourier-Laplace transforms of $\hat{\psi}_0(k,s)$, we
  obtain Eq.~(\ref{e.joint-eq}).

\bibliographystyle{apsrev} 

\begin{thebibliography}{45}
\expandafter\ifx\csname natexlab\endcsname\relax\def\natexlab#1{#1}\fi
\expandafter\ifx\csname bibnamefont\endcsname\relax
  \def\bibnamefont#1{#1}\fi
\expandafter\ifx\csname bibfnamefont\endcsname\relax
  \def\bibfnamefont#1{#1}\fi
\expandafter\ifx\csname citenamefont\endcsname\relax
  \fi
\expandafter\ifx\csname url\endcsname\relax
  \def\url#1{\texttt{#1}}\fi
\expandafter\ifx\csname urlprefix\endcsname\relax\fi
\providecommand{\bibinfo}[2]{#2}
\providecommand{\eprint}[2][]{\url{#2}}

\bibitem[1]{Caspi2000}
\bibinfo{author}{\bibfnamefont{A.}~\bibnamefont{Caspi}},
  \bibinfo{author}{\bibfnamefont{R.}~\bibnamefont{Granek}}, \bibnamefont{and}
  \bibinfo{author}{\bibfnamefont{M.}~\bibnamefont{Elbaum}},
  \bibinfo{journal}{Phys. Rev. Lett.} \textbf{\bibinfo{volume}{85}},
  \bibinfo{pages}{5655} (\bibinfo{year}{2000}).

\bibitem[2]{Golding2006}
\bibinfo{author}{\bibfnamefont{I.}~\bibnamefont{Golding}} \bibnamefont{and}
  \bibinfo{author}{\bibfnamefont{E.~C.} \bibnamefont{Cox}},
  \bibinfo{journal}{Phys. Rev. Lett.} \textbf{\bibinfo{volume}{96}},
  \bibinfo{pages}{098102} (\bibinfo{year}{2006}).

\bibitem[3]{Graneli2006}
\bibinfo{author}{\bibfnamefont{A.}~\bibnamefont{Gran\'{e}li}},
  \bibinfo{author}{\bibfnamefont{C.~C.} \bibnamefont{Yeykal}},
  \bibinfo{author}{\bibfnamefont{R.~B.} \bibnamefont{Robertson}},
  \bibnamefont{and} \bibinfo{author}{\bibfnamefont{E.~C.}
  \bibnamefont{Greene}}, \bibinfo{journal}{Proc. Natl. Acad. Sci. USA}
  \textbf{\bibinfo{volume}{103}}, \bibinfo{pages}{1221} (\bibinfo{year}{2006}).

\bibitem[4]{Weigel2011}
\bibinfo{author}{\bibfnamefont{A.}~\bibnamefont{Weigel}},
  \bibinfo{author}{\bibfnamefont{B.}~\bibnamefont{Simon}},
  \bibinfo{author}{\bibfnamefont{M.}~\bibnamefont{Tamkun}}, \bibnamefont{and}
  \bibinfo{author}{\bibfnamefont{D.}~\bibnamefont{Krapf}},
  \bibinfo{journal}{Proc. Natl. Acad. Sci. USA} \textbf{\bibinfo{volume}{108}},
  \bibinfo{pages}{6438} (\bibinfo{year}{2011}).

\bibitem[5]{Jeon2011}
\bibinfo{author}{\bibfnamefont{J.-H.} \bibnamefont{Jeon}},
  \bibinfo{author}{\bibfnamefont{V.}~\bibnamefont{Tejedor}},
  \bibinfo{author}{\bibfnamefont{S.}~\bibnamefont{Burov}},
  \bibinfo{author}{\bibfnamefont{E.}~\bibnamefont{Barkai}},
  \bibinfo{author}{\bibfnamefont{C.}~\bibnamefont{Selhuber-Unkel}},
  \bibinfo{author}{\bibfnamefont{K.}~\bibnamefont{Berg-S\o{}rensen}},
  \bibinfo{author}{\bibfnamefont{L.}~\bibnamefont{Oddershede}},
  \bibnamefont{and} \bibinfo{author}{\bibfnamefont{R.}~\bibnamefont{Metzler}},
  \bibinfo{journal}{Phys. Rev. Lett.} \textbf{\bibinfo{volume}{106}},
  \bibinfo{pages}{048103} (\bibinfo{year}{2011}).

\bibitem[6]{Weber2012}
\bibinfo{author}{\bibfnamefont{S.~C.} \bibnamefont{Weber}},
  \bibinfo{author}{\bibfnamefont{A.~J.} \bibnamefont{Spakowitz}},
  \bibnamefont{and} \bibinfo{author}{\bibfnamefont{J.~A.}
  \bibnamefont{Theriot}}, \bibinfo{journal}{Proc. Natl. Acad. Sci. USA}
  \textbf{\bibinfo{volume}{109}}, \bibinfo{pages}{7338} (\bibinfo{year}{2012}).

\bibitem[7]{Tabei2013}
\bibinfo{author}{\bibfnamefont{S.~A.} \bibnamefont{Tabei}},
  \bibinfo{author}{\bibfnamefont{S.}~\bibnamefont{Burov}},
  \bibinfo{author}{\bibfnamefont{H.~Y.} \bibnamefont{Kim}},
  \bibinfo{author}{\bibfnamefont{A.}~\bibnamefont{Kuznetsov}},
  \bibinfo{author}{\bibfnamefont{T.}~\bibnamefont{Huynh}},
  \bibinfo{author}{\bibfnamefont{J.}~\bibnamefont{Jureller}},
  \bibinfo{author}{\bibfnamefont{L.~H.} \bibnamefont{Philipson}},
  \bibinfo{author}{\bibfnamefont{A.~R.} \bibnamefont{Dinner}},
  \bibnamefont{and} \bibinfo{author}{\bibfnamefont{N.~F.}
  \bibnamefont{Scherer}}, \bibinfo{journal}{Proc. Natl. Acad. Sci. USA}
  \textbf{\bibinfo{volume}{110}}, \bibinfo{pages}{4911} (\bibinfo{year}{2013}).

\bibitem[8]{Magdziarz2009}
\bibinfo{author}{\bibfnamefont{M.}~\bibnamefont{Magdziarz}},
  \bibinfo{author}{\bibfnamefont{A.}~\bibnamefont{Weron}},
  \bibinfo{author}{\bibfnamefont{K.}~\bibnamefont{Burnecki}}, \bibnamefont{and}
  \bibinfo{author}{\bibfnamefont{J.}~\bibnamefont{Klafter}},
  \bibinfo{journal}{Phys. Rev. Lett.} \textbf{\bibinfo{volume}{103}},
  \bibinfo{pages}{180602} (\bibinfo{year}{2009}).

\bibitem[9]{Tejedor2010a}
\bibinfo{author}{\bibfnamefont{V.}~\bibnamefont{Tejedor}},
  \bibinfo{author}{\bibfnamefont{O.}~\bibnamefont{B{\'e}nichou}},
  \bibinfo{author}{\bibfnamefont{R.}~\bibnamefont{Voituriez}},
  \bibinfo{author}{\bibfnamefont{R.}~\bibnamefont{Jungmann}},
  \bibinfo{author}{\bibfnamefont{F.}~\bibnamefont{Simmel}},
  \bibinfo{author}{\bibfnamefont{C.}~\bibnamefont{Selhuber-Unkel}},
  \bibinfo{author}{\bibfnamefont{L.~B.} \bibnamefont{Oddershede}},
  \bibnamefont{and} \bibinfo{author}{\bibfnamefont{R.}~\bibnamefont{Metzler}},
  \bibinfo{journal}{Biophysical J.} \textbf{\bibinfo{volume}{98}},
  \bibinfo{pages}{1364} (\bibinfo{year}{2010}).

\bibitem[10]{Kepten2011}
\bibinfo{author}{\bibfnamefont{E.}~\bibnamefont{Kepten}},
  \bibinfo{author}{\bibfnamefont{I.}~\bibnamefont{Bronshtein}},
  \bibnamefont{and} \bibinfo{author}{\bibfnamefont{Y.}~\bibnamefont{Garini}},
  \bibinfo{journal}{Phys. Rev. E} \textbf{\bibinfo{volume}{83}},
  \bibinfo{pages}{041919} (\bibinfo{year}{2011}).

\bibitem[11]{Magdziarz2011}
\bibinfo{author}{\bibfnamefont{M.}~\bibnamefont{Magdziarz}} \bibnamefont{and}
  \bibinfo{author}{\bibfnamefont{A.}~\bibnamefont{Weron}},
  \bibinfo{journal}{Phys. Rev. E} \textbf{\bibinfo{volume}{84}},
  \bibinfo{pages}{051138} (\bibinfo{year}{2011}).

\bibitem[12]{Meroz2013}
\bibinfo{author}{\bibfnamefont{Y.}~\bibnamefont{Meroz}},
  \bibinfo{author}{\bibfnamefont{I.~M.} \bibnamefont{Sokolov}},
  \bibnamefont{and} \bibinfo{author}{\bibfnamefont{J.}~\bibnamefont{Klafter}},
  \bibinfo{journal}{Phys. Rev. Lett.} \textbf{\bibinfo{volume}{110}},
  \bibinfo{pages}{090601} (\bibinfo{year}{2013}).

\bibitem[13]{Lubelski2008}
\bibinfo{author}{\bibfnamefont{A.}~\bibnamefont{Lubelski}},
  \bibinfo{author}{\bibfnamefont{I.~M.} \bibnamefont{Sokolov}},
  \bibnamefont{and} \bibinfo{author}{\bibfnamefont{J.}~\bibnamefont{Klafter}},
  \bibinfo{journal}{Phys. Rev. Lett.} \textbf{\bibinfo{volume}{100}},
  \bibinfo{pages}{250602} (\bibinfo{year}{2008}).

\bibitem[14]{He2008}
\bibinfo{author}{\bibfnamefont{Y.}~\bibnamefont{He}},
  \bibinfo{author}{\bibfnamefont{S.}~\bibnamefont{Burov}},
  \bibinfo{author}{\bibfnamefont{R.}~\bibnamefont{Metzler}}, \bibnamefont{and}
  \bibinfo{author}{\bibfnamefont{E.}~\bibnamefont{Barkai}},
  \bibinfo{journal}{Phys. Rev. Lett.} \textbf{\bibinfo{volume}{101}},
  \bibinfo{pages}{058101} (\bibinfo{year}{2008}).

\bibitem[15]{Miyaguchi2011a}
\bibinfo{author}{\bibfnamefont{T.}~\bibnamefont{Miyaguchi}} \bibnamefont{and}
  \bibinfo{author}{\bibfnamefont{T.}~\bibnamefont{Akimoto}},
  \bibinfo{journal}{Phys. Rev. E} \textbf{\bibinfo{volume}{83}},
  \bibinfo{pages}{062101} (\bibinfo{year}{2011}).

\bibitem[16]{Miyaguchi2011}
\bibinfo{author}{\bibfnamefont{T.}~\bibnamefont{Miyaguchi}} \bibnamefont{and}
  \bibinfo{author}{\bibfnamefont{T.}~\bibnamefont{Akimoto}},
  \bibinfo{journal}{Phys. Rev. E} \textbf{\bibinfo{volume}{83}},
  \bibinfo{pages}{031926} (\bibinfo{year}{2011}).

\bibitem[17]{Akimoto2010}
\bibinfo{author}{\bibfnamefont{T.}~\bibnamefont{Akimoto}} \bibnamefont{and}
  \bibinfo{author}{\bibfnamefont{T.}~\bibnamefont{Miyaguchi}},
  \bibinfo{journal}{Phys. Rev. E} \textbf{\bibinfo{volume}{82}},
  \bibinfo{pages}{030102(R)} (\bibinfo{year}{2010}).

\bibitem[18]{bouchaud90}
\bibinfo{author}{\bibfnamefont{J.}~\bibnamefont{Bouchaud}} \bibnamefont{and}
  \bibinfo{author}{\bibfnamefont{A.}~\bibnamefont{Georges}},
  \bibinfo{journal}{Phys. Rep.} \textbf{\bibinfo{volume}{195}},
  \bibinfo{pages}{127} (\bibinfo{year}{1990}).

\bibitem[19]{Akimoto2010a}
\bibinfo{author}{\bibfnamefont{T.}~\bibnamefont{Akimoto}} \bibnamefont{and}
  \bibinfo{author}{\bibfnamefont{Y.}~\bibnamefont{Aizawa}},
  \bibinfo{journal}{Chaos} \textbf{\bibinfo{volume}{20}},
  \bibinfo{pages}{033110} (\bibinfo{year}{2011}).

\bibitem[20]{Aaronson1997}
\bibinfo{author}{\bibfnamefont{J.}~\bibnamefont{Aaronson}},
  \emph{\bibinfo{title}{An Introduction to Infinite Ergodic Theory}}
  (\bibinfo{publisher}{American Mathematical Society},
  \bibinfo{address}{Providence}, \bibinfo{year}{1997}).

\bibitem[21]{Brok2003}
\bibinfo{author}{\bibfnamefont{X.}~\bibnamefont{Brokmann}},
  \bibinfo{author}{\bibfnamefont{J.-P.} \bibnamefont{Hermier}},
  \bibinfo{author}{\bibfnamefont{G.}~\bibnamefont{Messin}},
  \bibinfo{author}{\bibfnamefont{P.}~\bibnamefont{Desbiolles}},
  \bibinfo{author}{\bibfnamefont{J.-P.} \bibnamefont{Bouchaud}},
  \bibnamefont{and} \bibinfo{author}{\bibfnamefont{M.}~\bibnamefont{Dahan}},
  \bibinfo{journal}{Phys. Rev. Lett.} \textbf{\bibinfo{volume}{90}},
  \bibinfo{pages}{120601} (\bibinfo{year}{2003}).

\bibitem[22]{Stefani2009}
\bibinfo{author}{\bibfnamefont{F.~D.} \bibnamefont{Stefani}},
  \bibinfo{author}{\bibfnamefont{J.~P.} \bibnamefont{Hoogenboom}},
  \bibnamefont{and} \bibinfo{author}{\bibfnamefont{E.}~\bibnamefont{Barkai}},
  \bibinfo{journal}{Phys. Today} \textbf{\bibinfo{volume}{62}},
  \bibinfo{pages}{34} (\bibinfo{year}{2009}).

\bibitem[23]{Akimoto2013a}
\bibinfo{author}{\bibfnamefont{T.}~\bibnamefont{Akimoto}} \bibnamefont{and}
  \bibinfo{author}{\bibfnamefont{T.}~\bibnamefont{Miyaguchi}},
  \bibinfo{journal}{Phys. Rev. E} \textbf{\bibinfo{volume}{87}},
  \bibinfo{pages}{062134} (\bibinfo{year}{2013}).

\bibitem[24]{Klafter1987}
\bibinfo{author}{\bibfnamefont{J.}~\bibnamefont{Klafter}},
  \bibinfo{author}{\bibfnamefont{A.}~\bibnamefont{Blumen}}, \bibnamefont{and}
  \bibinfo{author}{\bibfnamefont{M.~F.} \bibnamefont{Shlesinger}},
  \bibinfo{journal}{Phys. Rev. A} \textbf{\bibinfo{volume}{35}},
  \bibinfo{pages}{3081} (\bibinfo{year}{1987}).

\bibitem[25]{Magdziarz2012}
\bibinfo{author}{\bibfnamefont{M.}~\bibnamefont{Magdziarz}},
  \bibinfo{author}{\bibfnamefont{W.}~\bibnamefont{Szczotka}}, \bibnamefont{and}
  \bibinfo{author}{\bibfnamefont{P.}~\bibnamefont{{\.Z}ebrowski}},
  \bibinfo{journal}{J. Stat. Phys.} \textbf{\bibinfo{volume}{147}},
  \bibinfo{pages}{74} (\bibinfo{year}{2012}).

\bibitem[26]{Liu2013}
\bibinfo{author}{\bibfnamefont{J.}~\bibnamefont{Liu}} \bibnamefont{and}
  \bibinfo{author}{\bibfnamefont{J.-D.} \bibnamefont{Bao}},
  \bibinfo{journal}{Physica A} \textbf{\bibinfo{volume}{392}},
  \bibinfo{pages}{612} (\bibinfo{year}{2013}).

\bibitem[27]{Shlesinger1987}
\bibinfo{author}{\bibfnamefont{M.~F.} \bibnamefont{Shlesinger}},
  \bibinfo{author}{\bibfnamefont{B.~J.} \bibnamefont{West}}, \bibnamefont{and}
  \bibinfo{author}{\bibfnamefont{J.}~\bibnamefont{Klafter}},
  \bibinfo{journal}{Phys. Rev. Lett.} \textbf{\bibinfo{volume}{58}},
  \bibinfo{pages}{1100} (\bibinfo{year}{1987}).

\bibitem[28]{Akimoto2012}
\bibinfo{author}{\bibfnamefont{T.}~\bibnamefont{Akimoto}},
  \bibinfo{journal}{Phys. Rev. Lett.} \textbf{\bibinfo{volume}{108}},
  \bibinfo{pages}{164101} (\bibinfo{year}{2012}).

\bibitem[29]{Froemberg2013}
\bibinfo{author}{\bibfnamefont{D.}~\bibnamefont{Froemberg}} \bibnamefont{and}
  \bibinfo{author}{\bibfnamefont{E.}~\bibnamefont{Barkai}},
  \bibinfo{journal}{Phys. Rev. E} \textbf{\bibinfo{volume}{87}},
  \bibinfo{pages}{030104} (\bibinfo{year}{2013}{\natexlab{a}}).

\bibitem[30]{Barkai2014}
\bibinfo{author}{\bibfnamefont{E.}~\bibnamefont{Barkai}},
  \bibinfo{author}{\bibfnamefont{E.}~\bibnamefont{Aghion}}, \bibnamefont{and}
  \bibinfo{author}{\bibfnamefont{D.}~\bibnamefont{Kessler}},
  \bibinfo{journal}{Phys. Rev. X} \textbf{\bibinfo{volume}{4}},
  \bibinfo{pages}{021036} (\bibinfo{year}{2014}).

\bibitem[31]{Meerschaert2006}
\bibinfo{author}{\bibfnamefont{M.~M.} \bibnamefont{Meerschaert}}
  \bibnamefont{and} \bibinfo{author}{\bibfnamefont{E.}~\bibnamefont{Scalas}},
  \bibinfo{journal}{Physica A} \textbf{\bibinfo{volume}{370}},
  \bibinfo{pages}{114} (\bibinfo{year}{2006}).

\bibitem[32]{Corral2006}
\bibinfo{author}{\bibfnamefont{A.}~\bibnamefont{Corral}},
  \bibinfo{journal}{Phys. Rev. Lett.} \textbf{\bibinfo{volume}{97}},
  \bibinfo{pages}{178501} (\bibinfo{year}{2006}).

\bibitem[33]{Lippiello2013}
\bibinfo{author}{\bibfnamefont{E.}~\bibnamefont{Lippiello}},
  \bibinfo{author}{\bibfnamefont{C.}~\bibnamefont{Godano}}, \bibnamefont{and}
  \bibinfo{author}{\bibfnamefont{L.}~\bibnamefont{de~Arcangelis}},
  \bibinfo{journal}{Europhys. Lett.} \textbf{\bibinfo{volume}{102}},
  \bibinfo{pages}{59002} (\bibinfo{year}{2013}).

\bibitem[34]{Cox}
\bibinfo{author}{\bibfnamefont{D.~R.} \bibnamefont{Cox}},
  \emph{\bibinfo{title}{Renewal theory}} (\bibinfo{publisher}{Methuen},
  \bibinfo{address}{London}, \bibinfo{year}{1962}).

\bibitem[35]{Shlesinger1982}
\bibinfo{author}{\bibfnamefont{M.}~\bibnamefont{Shlesinger}},
  \bibinfo{author}{\bibfnamefont{J.}~\bibnamefont{Klafter}}, \bibnamefont{and}
  \bibinfo{author}{\bibfnamefont{Y.}~\bibnamefont{Wong}}, \bibinfo{journal}{J.
  Stat. Phys.} \textbf{\bibinfo{volume}{27}}, \bibinfo{pages}{499}
  (\bibinfo{year}{1982}).

\bibitem[36]{Barkai2003}
\bibinfo{author}{\bibfnamefont{E.}~\bibnamefont{Barkai}},
  \bibinfo{journal}{Phys. Rev. Lett.} \textbf{\bibinfo{volume}{90}},
  \bibinfo{pages}{104101} (\bibinfo{year}{2003}).

\bibitem[37]{Schulz2013}
\bibinfo{author}{\bibfnamefont{J.~H.~P.} \bibnamefont{Schulz}},
  \bibinfo{author}{\bibfnamefont{E.}~\bibnamefont{Barkai}}, \bibnamefont{and}
  \bibinfo{author}{\bibfnamefont{R.}~\bibnamefont{Metzler}},
  \bibinfo{journal}{Phys. Rev. Lett.} \textbf{\bibinfo{volume}{110}},
  \bibinfo{pages}{020602} (\bibinfo{year}{2013}).

\bibitem[38]{Akimoto2013c}
\bibinfo{author}{\bibfnamefont{T.}~\bibnamefont{Akimoto}},
  \bibinfo{author}{\bibfnamefont{S.}~\bibnamefont{Shinkai}}, \bibnamefont{and}
  \bibinfo{author}{\bibfnamefont{Y.}~\bibnamefont{Aizawa}},
  \bibinfo{journal}{arxiv:1310.4055}.

\bibitem[39]{Zumofen1993}
\bibinfo{author}{\bibfnamefont{G.}~\bibnamefont{Zumofen}} \bibnamefont{and}
  \bibinfo{author}{\bibfnamefont{J.}~\bibnamefont{Klafter}},
  \bibinfo{journal}{Physica D} \textbf{\bibinfo{volume}{69}},
  \bibinfo{pages}{436} (\bibinfo{year}{1993}).

\bibitem[40]{Godec2013}
\bibinfo{author}{\bibfnamefont{A.}~\bibnamefont{Godec}} \bibnamefont{and}
  \bibinfo{author}{\bibfnamefont{R.}~\bibnamefont{Metzler}},
  \bibinfo{journal}{Phys. Rev. Lett.} \textbf{\bibinfo{volume}{110}},
  \bibinfo{pages}{020603} (\bibinfo{year}{2013}).

\bibitem[41]{Froemberg2013b}
\bibinfo{author}{\bibfnamefont{D.}~\bibnamefont{Froemberg}} \bibnamefont{and}
  \bibinfo{author}{\bibfnamefont{E.}~\bibnamefont{Barkai}},
  \bibinfo{journal}{Eur. Phys. J. B} \textbf{\bibinfo{volume}{86}},
  \bibinfo{pages}{331} (\bibinfo{year}{2013}{\natexlab{b}}).

\bibitem[42]{Akimoto2007}
\bibinfo{author}{\bibfnamefont{T.}~\bibnamefont{Akimoto}} \bibnamefont{and}
  \bibinfo{author}{\bibfnamefont{Y.}~\bibnamefont{Aizawa}}, \bibinfo{journal}{J
  Korean Phys. Soc.} \textbf{\bibinfo{volume}{50}}, \bibinfo{pages}{254}
  (\bibinfo{year}{2007}).

\bibitem[43]{Miyaguchi2013}
\bibinfo{author}{\bibfnamefont{T.}~\bibnamefont{Miyaguchi}} \bibnamefont{and}
  \bibinfo{author}{\bibfnamefont{T.}~\bibnamefont{Akimoto}},
  \bibinfo{journal}{Phys. Rev. E} \textbf{\bibinfo{volume}{87}},
  \bibinfo{pages}{032130} (\bibinfo{year}{2013}).

\bibitem[44]{Akimoto2011}
\bibinfo{author}{\bibfnamefont{T.}~\bibnamefont{Akimoto}},
  \bibinfo{author}{\bibfnamefont{E.}~\bibnamefont{Yamamoto}},
  \bibinfo{author}{\bibfnamefont{K.}~\bibnamefont{Yasuoka}},
  \bibinfo{author}{\bibfnamefont{Y.}~\bibnamefont{Hirano}}, \bibnamefont{and}
  \bibinfo{author}{\bibfnamefont{M.}~\bibnamefont{Yasui}},
  \bibinfo{journal}{Phys. Rev. Lett.} \textbf{\bibinfo{volume}{107}},
  \bibinfo{pages}{178103} (\bibinfo{year}{2011}).

\bibitem[45]{Uneyama2012}
\bibinfo{author}{\bibfnamefont{T.}~\bibnamefont{Uneyama}},
  \bibinfo{author}{\bibfnamefont{T.}~\bibnamefont{Akimoto}}, \bibnamefont{and}
  \bibinfo{author}{\bibfnamefont{T.}~\bibnamefont{Miyaguchi}},
  \bibinfo{journal}{J. Chem. Phys.} \textbf{\bibinfo{volume}{137}},
  \bibinfo{pages}{114903} (\bibinfo{year}{2012}).

\end{thebibliography}

\end {document}